\documentclass[aps,prl,twocolumn,superscriptaddress,amsfont,graphicx,nofootinbib,preprintnumbers]{revtex4-1}
\usepackage{color,graphicx,epsfig}
\usepackage{ifpdf}
\usepackage{amsmath}
\usepackage{bm}
\usepackage{color}
\usepackage[english]{babel}
\usepackage{graphicx}
\usepackage{amsfonts}
\usepackage{amssymb}
\usepackage{hyperref}
\usepackage{slashed}

\definecolor{nicered}{rgb}{0.7,0.1,0.1}
\definecolor{nicegreen}{rgb}{0.1,0.5,0.1}
\hypersetup{colorlinks,citecolor= nicegreen,linkcolor= nicered}

\graphicspath{{./figs/}}
\begin{document}

\def\Oxford{Rudolf Peierls Centre for Theoretical Physics, University of Oxford OX1 3NP Oxford, United Kingdom}
\def\CERN{CERN, Theoretical Physics Department, CH-1211 Geneva 23, Switzerland}

\preprint{OUTP-17-06P, CERN-TH-2017-104}

\title{\boldmath On Light Resonance Interpretations of the $B$ Decay Anomalies}

\author{Fady Bishara}
\email[Electronic address:]{fady.bishara@physics.ox.ac.uk} 
\affiliation{\Oxford}
\author{Ulrich Haisch}
\email[Electronic address:]{ulrich.haisch@physics.ox.ac.uk} 
\affiliation{\Oxford}
\affiliation{\CERN}
\author{Pier Francesco Monni} 
\email[Electronic address:]{pier.monni@cern.ch} 
\affiliation{\CERN}

\begin{abstract} 
We sketch a novel method to search for light di-leptonic resonances by exploiting precision measurements of Drell-Yan production. Motivated by the recent hints of lepton flavour universality violation in $B \to K^{\ast} \ell^+ \ell^-$, we illustrate our proposal by studying the case of spin-1 resonances that couple to muons and have masses in the range of a few GeV. We show that the existing LHC data on $pp \to Z/\gamma^\ast \to \mu^+ \mu^-$ put non-trivial constraints on light di-muon resonance interpretations of  $B$ decay anomalies in a model-independent fashion. The impact of our proposal on the long-standing discrepancy in the anomalous magnetic moment of the muon is also briefly discussed. 
\end{abstract}

\maketitle

{\bf Introduction.} In the last four years several  anomalies have been observed in rare semi-leptonic $B$ decays governed by $b \to s \ell^+ \ell^-$ transitions. Specifically, deviations from the standard model (SM) expectations in the angular observable $P_5^\prime$ in $B \to K^\ast \mu^+ \mu^-$~\cite{Aaij:2013qta,Aaij:2015oid,Wehle:2016yoi,ATLAS-CONF-2017-023}, the branching ratios of $B^+ \to K^{(\ast ) +} \mu^+ \mu^-$~\cite{Aaij:2014pli}, $B \to K^{(\ast )} \mu^+ \mu^-$~\cite{Aaij:2014pli,Aaij:2016flj} and $B_s \to \phi \mu^+ \mu^-$~\cite{Aaij:2015esa} as well as the ratio~$R_K$ of di-muon to di-electron rates in $B^+ \to K^+ \ell^+ \ell^-$~\cite{Aaij:2014ora} have been reported. The recent measurement of the ratio~$R_{K^\ast}$ of di-muon to di-electron rates in $B \to K^{\ast} \ell^+ \ell^-$ has added to the list of anomalies~\cite{RKstar} and has, accordingly, caught the attention of the theory community~\cite{Capdevila:2017bsm,Altmannshofer:2017yso,DAmico:2017mtc,Hiller:2017bzc,Geng:2017svp,Ciuchini:2017mik,Celis:2017doq,Becirevic:2017jtw,Cai:2017wry,Kamenik:2017tnu,Sala:2017ihs,Ghosh:2017ber,DiChiara:2017cjq,Alok:2017jaf,Alok:2017sui,Wang:2017mrd,Greljo:2017vvb,Bonilla:2017lsq,Feruglio:2017rjo}.

Although each deviation by itself is not  statistically significant, and the angular observables and branching ratios are afflicted by hadronic uncertainties that obscure the interpretation and significance of the anomalies, it is quite intriguing that the deviations seen in the theoretically clean lepton-universality ratios $R_K$ and $R_{K^\ast}$ might be part of a coherent picture~\cite{Capdevila:2017bsm,Altmannshofer:2017yso,DAmico:2017mtc,Geng:2017svp,Ciuchini:2017mik} involving  new physics in the $b \to s \mu^+ \mu^-$ transitions in the form of the two dimension-six operators $Q_9 = (\bar s_L \gamma_\alpha b_L)( \bar \mu \gamma^\alpha \mu)$ and $Q_{10} =( \bar s_L \gamma_\alpha b_L)( \bar \mu \gamma^\alpha \gamma_5 \mu)$.

The most popular new-physics interpretations that can accommodate the $b \to s \ell^+ \ell^-$ anomalies involve new heavy degrees of freedom such as $Z^\prime$ bosons or lepto-quarks (see~e.g.~\cite{DAmico:2017mtc} and references therein). Solutions that involve a new light resonance have instead received  less attention~\cite{Sala:2017ihs,Ghosh:2017ber,Fuyuto:2015gmk,Datta:2017pfz,Alok:2017sui},\footnote{The possibility that a light resonance could be responsible for the anomaly in $P_5^\prime$ was  mentioned by Amarjit Soni at 50th Rencontres de Moriond EW 2015, and subsequently re-emphasised to one of the authors by Brian Batell in a private conversation.} although they might offer an explanation of the long-standing discrepancy~(cf.~\cite{gmtwo}) in the  anomalous magnetic moment of the muon $a_\mu= \big ((g-2)/2 \big )_\mu$. In fact, it has been shown very recently~\cite{Sala:2017ihs} that a spin-1 resonance with  a mass of a  around $2.5 \, {\rm GeV}$ and a large invisible branching ratio can  simultaneously explain  both the flavour  anomalies and $a_\mu$  while evading various other constraints, if the  couplings of the mediator to fermions are dialed correctly.

In this letter, we point out that  light resonance interpretations of the $b \to s \ell^+ \ell^-$ anomalies can be tested and constrained through precision studies of Drell-Yan~(DY) production.\footnote{Constraints on  heavy di-lepton resonance interpretations of the $b \to s \ell^+ \ell^-$ tensions using present and future $pp \to \ell^+ \ell^-$ data have been very recently derived in~\cite{Greljo:2017vvb}.} Our finding is based on the simple observation that  final state radiation (FSR) of a light di-leptonic resonance in $pp \to Z/\gamma^\ast \to \ell^+ \ell^-$ will lead to observable modifications of the kinematic distributions of the~$\ell^+ \ell^-$ system. We will illustrate this general idea by setting limits on the muon couplings of spin-1 resonances with masses in the GeV range by exploiting existing LHC data on the di-muon invariant mass~$m_{\mu \mu}$  close to the $Z$ peak. The impact  of  this novel model-independent bounds on spin-1 mediator interpretations of  the  anomalies observed in rare semi-leptonic $B$ decays as well as $a_\mu$  will be discussed in some detail. 

{\bf Simplified model.}  Following~\cite{Sala:2017ihs} we consider a simplified model valid at GeV energies which, besides the~SM particles, contains a colourless spin-1 mediator $V$ with mass $m_V$ and a SM singlet Dirac fermion $\chi$ of mass $m_\chi$.  The interactions of $V$ relevant for the further discussion are
\begin{equation} \label{eq:simplifiedmodel}
\begin{split}
{\cal L} & \supset \left ( g_L^{sb} \hspace{0.35mm} \bar s_L \slashed{V} b_L + {\rm h.c.} \right ) +  \bar \mu  \left  ( g_V^\mu - g_A^\mu   \hspace{0.25mm} \gamma_5 \right ) \slashed{V} \mu \\[2mm] & \phantom{xx} + g_V^\chi  \hspace{0.25mm} \bar \chi \slashed{V} \chi \,, 
\end{split}
\end{equation}
where, for concreteness, the couplings $g_L^{sb}$, $g_V^\mu$, $g_A^\mu$ and~$g_V^\chi$ are taken to be real, $\slashed{V} =  \gamma_\alpha V^\alpha$ and  the subscript $L$ denotes left-handed fermionic fields. In what follows we will assume that $g_L^{sb}$, $g_V^\mu$, $g_A^\mu$ and~$g_V^\chi$ encode all couplings between the new spin-1 state $V$ and fermions, and we will not specify an explicit ultraviolet completion that gives rise to them. We however add that the interactions~(\ref{eq:simplifiedmodel}) can emerge in various ways such as in vector-like fermion extensions or by considering an effective approach where all~$V$ couplings are generated via higher-dimensional operators~(see~e.g.~\cite{Fox:2011qd,Carone:2013uh}).

As demonstrated in~\cite{Sala:2017ihs}, to qualitatively reproduce the $P_5^\prime$, $R_K$, and $R_{K^\ast}$ anomalies, the mass of the new spin-1 mediator is constrained to lie in the range of about~$[2,3] \, {\rm GeV}$ and its total decay width has to satisfy $\Gamma_V/m_V \gtrsim 10 \%$. The total width requirement implies that $m_\chi < m_V/2$ and $g_\chi \gtrsim 2$. Consequently,~$V$ predominantly decays invisibly with a branching ratio ${\rm Br} (V \to \bar \chi \chi) \simeq 1$. The existence of a di-muon resonance with these properties cannot be excluded because of the large hadronic uncertainties of the SM prediction for $B \to K^{(\ast)} \mu^+ \mu^-$ in the $m_{\mu \mu}^2 \gtrsim 6 \, {\rm GeV}^2$ region~(cf.~\cite{Khodjamirian:2010vf,Lyon:2014hpa}) and the unknown interference pattern between the $J/\psi$ and the SM short-distance contribution. By choosing the couplings in~(\ref{eq:simplifiedmodel}) to be $g_L^{sb} \simeq -10^{-8}$, $g_V^\mu \simeq 0.1$ and $g_A^\mu \simeq -0.44 \hspace{0.25mm} g_V^\mu$, the discussed simplified model then does not only provide a solution to the flavour anomalies but also ameliorates the discrepancy observed in $a_\mu$. Other constraints that arise from $B_s$--$\bar B_s$ mixing, searches for $B \to K^{(\ast)} + {\rm invisible}$~\cite{delAmoSanchez:2010bk,Lutz:2013ftz,Lees:2013kla,Grygier:2017tzo}, $B_s \to \mu^+ \mu^-$~\cite{CMS:2014xfa,Aaij:2017vad} and $Z \to 4 \mu$~\cite{CMS:2012bw,Aad:2014wra} as well as the precision measurements of $Z\mu \bar \mu$ couplings~\cite{ALEPH:2005ab} are all satisfied for this choice of parameters.

\begin{figure}[t!]
\begin{center}
\includegraphics[width=0.8 \columnwidth]{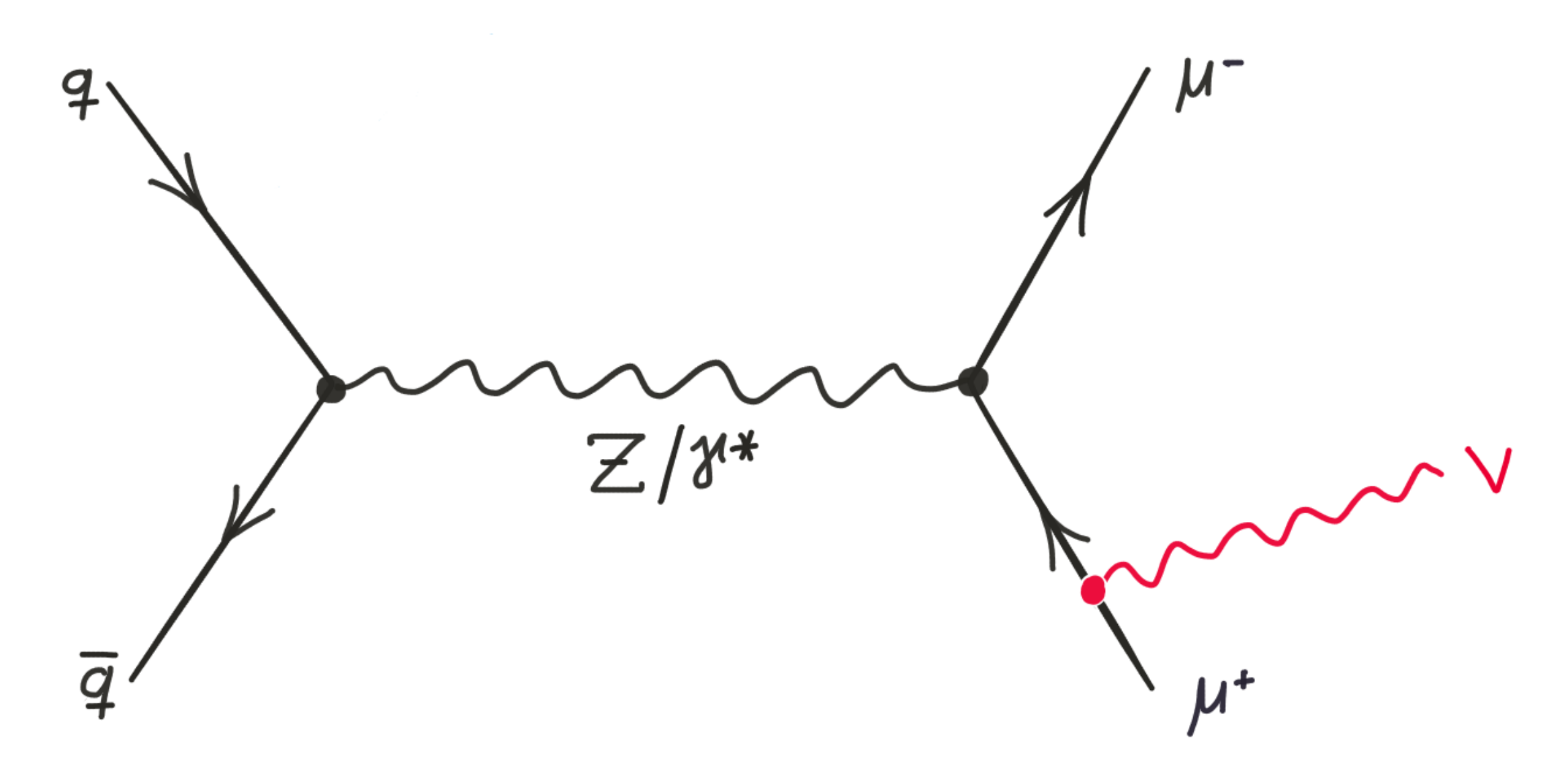}
\caption{An example of a Feynman diagram with FSR of the light mediator $V$ in the DY process $pp \to Z/\gamma^\ast \to \mu^+ \mu^-$.} 
\label{fig:FSR}
\end{center}
\end{figure}

{\bf $\bm{Z}$-boson line shape.}  We now consider the di-muon invariant mass spectrum as measured in DY production and study its distortions due to FSR of a light spin-1 resonance $V$. A representative diagram that contributes to $pp \to Z/\gamma^\ast \to \mu^+ \mu^- + V$ is shown in Figure~\ref{fig:FSR}. We calculate the $m_{\mu \mu}$ spectra with {\tt MadGraph5\_aMC@NLO}~\cite{Alwall:2014hca} and {\tt NNPDF23\_lo\_as\_0130\_qed} parton distribution functions~\cite{Ball:2013hta}, employing the {\tt DMsimp} implementation~\cite{Backovic:2015soa} of the $V \mu \bar \mu$ and $V \chi \bar \chi$ couplings in~(\ref{eq:simplifiedmodel}). The fiducial phase space in our Monte~Carlo simulations is defined by requiring that the muon transverse momenta satisfy $p_{T,\mu} > 25 \, {\rm GeV}$, the muon pseudorapidities obey $|\eta_\mu| < 2.5$, and that $m_{\mu \mu}$ falls into the range $[66, 116] \, {\rm GeV}$.

\begin{figure}[t!]
\begin{center}
\includegraphics[width=0.975 \columnwidth]{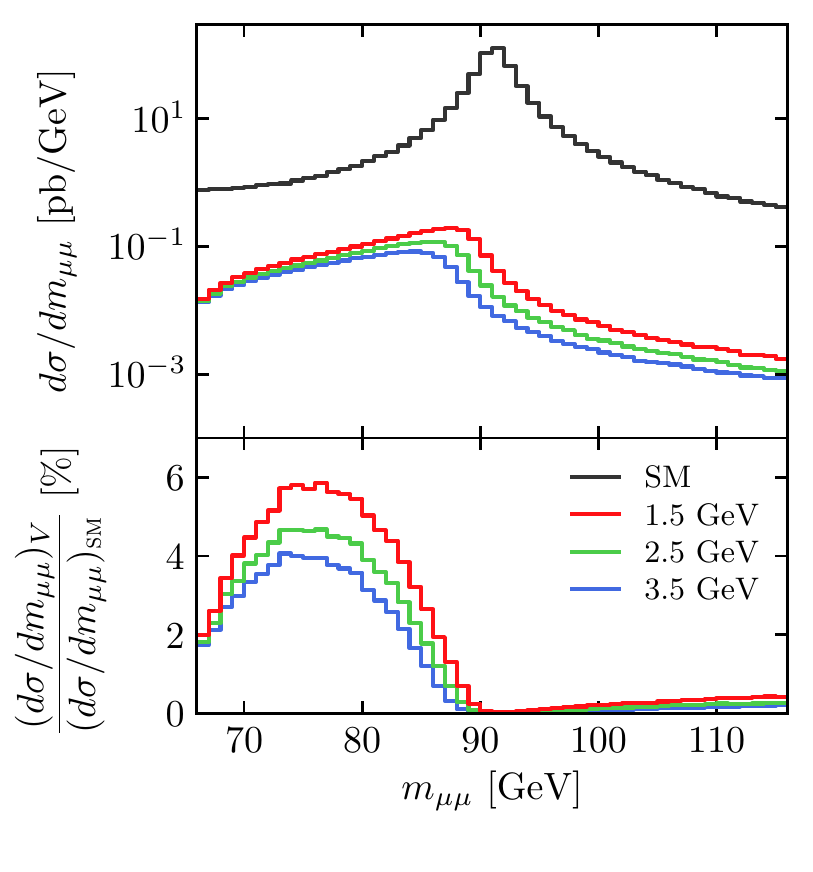}
\caption{Di-muon invariant mass  distributions. The black  curve represents the SM prediction, while the coloured curves correspond to three different benchmark models with varying spin-1 resonance mass. For further details see text.} 
\label{fig:lineshape}
\end{center}
\end{figure}

In Figure~\ref{fig:lineshape} we present our results for the di-muon invariant mass spectra for $pp$ collisions at a centre-of-mass energy of $\sqrt{s} = 13 \, {\rm TeV}$. All predictions are obtained at leading order in QCD. The three coloured curves correspond to $g_V^\mu = 0.1$ and $g_A^\mu = -4.4 \cdot 10^{-2}$ and mediator masses of $1.5\, {\rm GeV}$, $2.5 \, {\rm GeV}$ and $3.5 \, {\rm GeV}$, respectively.
For comparison, the SM prediction for the $Z$-boson line shape is depicted by the black curve. One observes that FSR of the spin-1 resonance leads to a pronounced radiation tail below $m_Z \simeq 91.2 \, {\rm GeV}$. This amounts to a relative correction to the SM $Z$-boson line shape of around 4\% to 6\% at $m_{\mu \mu} \simeq 75 \, {\rm GeV}$. 

DY processes are a cornerstone of the SM physics programme at the LHC~(see~e.g.~\cite{Aaboud:2016btc,Aaboud:2016zpd,Aaboud:2017hbk,CMS-PAS-SMP-15-002,CMS-PAS-SMP-15-010,CMS-PAS-SMP-16-009} for recent ATLAS and CMS analyses) and a detailed understanding of the $Z$-boson line shape is a  prerequisite for a precision measurement of the $W$-boson mass at the LHC~\cite{Aaboud:2017svj}. Given its importance, a lot of effort has gone into measuring the~$m_{\mu \mu}$ spectrum in the $Z$-peak region at the LHC and the experimental uncertainties have reached the few-percent level, making  the $Z$-boson line shape a powerful observable to search for GeV-mass di-muon spin-1 states. 

In our study we consider the ratio of the data to the SM prediction to perform a $\chi^2$ fit. In Figure~2 of~\cite{Aaboud:2016zpd}, the ATLAS collaboration provides the ratio of experimental data to the state-of-the-art theory prediction for the $m_{\mu \mu}$ line shape in the fiducial volume defined above. Assuming that the data is SM-like, we compute this ratio for different new-physics scenarios and perform a $\chi^2$ analysis. Radiative corrections of QCD and electroweak nature do not affect the ratio and are therefore neglected in the following. The ATLAS analysis is based on $3.2 \, {\rm fb}^{-1}$ integrated luminosity at $\sqrt{s} = 13 \, {\rm TeV}$. The experimental statistical and systematic uncertainties are in the range of 1\% to 2\%  and they are added in quadrature  in our fit. The bin-to-bin correlations are neglected.

\begin{figure}[t!]
\begin{center}
\includegraphics[width=0.975 \columnwidth]{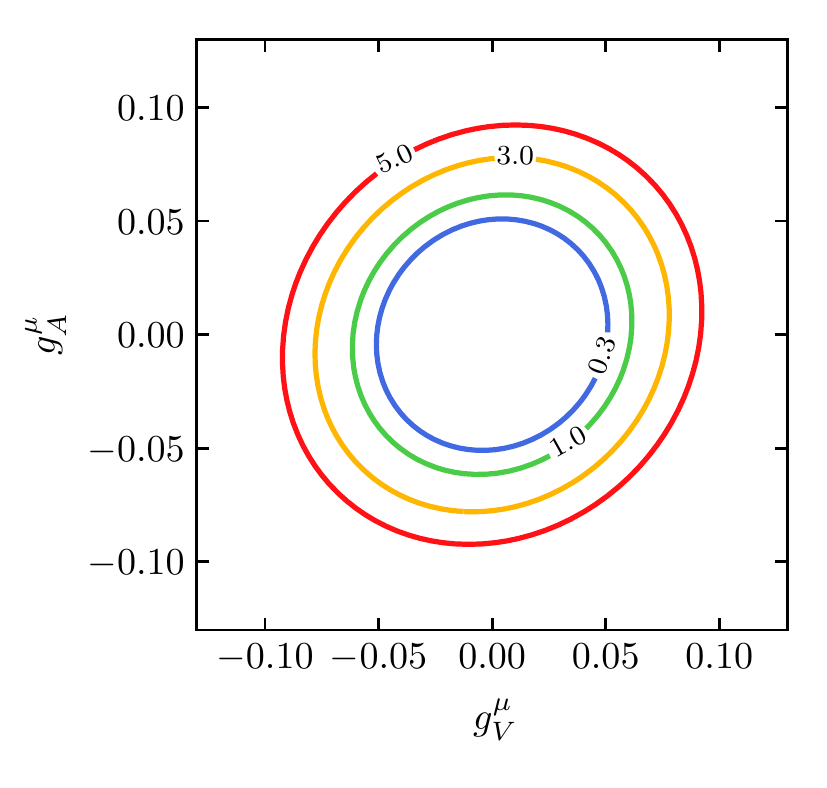}
\vspace{-4mm}
\caption{Constraints in the $g_V^\mu \hspace{0.25mm}$--$\hspace{0.25mm} g_A^\mu$ plane arising from DY production for different di-muon spin-1 resonance masses in units of GeV. The contours correspond to $\Delta\chi^2 = 5.99$. See text for additional details.}
\label{fig:modelindependent}
\end{center}
\end{figure}

In Figure~\ref{fig:modelindependent} we show the $\Delta \chi^2 =  5.99$ contours (corresponding to a $95\%$ confidence level (CL) for a Gaussian distribution) in the $g_V^\mu \hspace{0.25mm}$--$\hspace{0.25mm} g_A^\mu$ plane that follow from~our~$\chi^2$ analysis for different values of $m_V$. The parameter space outside the lines is disfavoured for each individual mass.  We find that for $m_V \in [1,5] \, {\rm GeV}$ the obtained 95\%~CL bounds can be approximately described by the inequality
\begin{equation} \label{eq:masterformula}
\sqrt{\left (g_V^\mu \right )^2 + \left (g_A^\mu \right )^2} \lesssim 5.6 \cdot 10^{-2} \, \left ( 1  + 0.13 \, \frac{m_V}{1 \, {\rm GeV}} \right ) \,.
\end{equation}
This approximate formula can be used to quickly assess the sensitivity of existing DY measurements on the coupling strength of GeV-mass di-muon spin-1 states. 

\begin{figure}[t!]
\begin{center}
\includegraphics[width=0.975 \columnwidth]{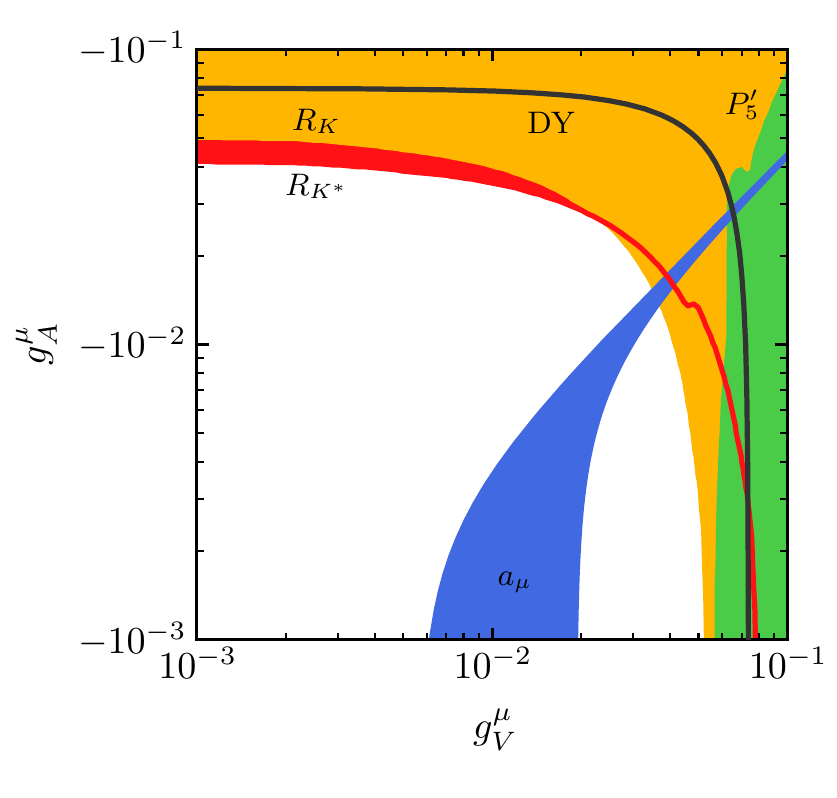}

\vspace{-4mm}

\includegraphics[width=0.975 \columnwidth]{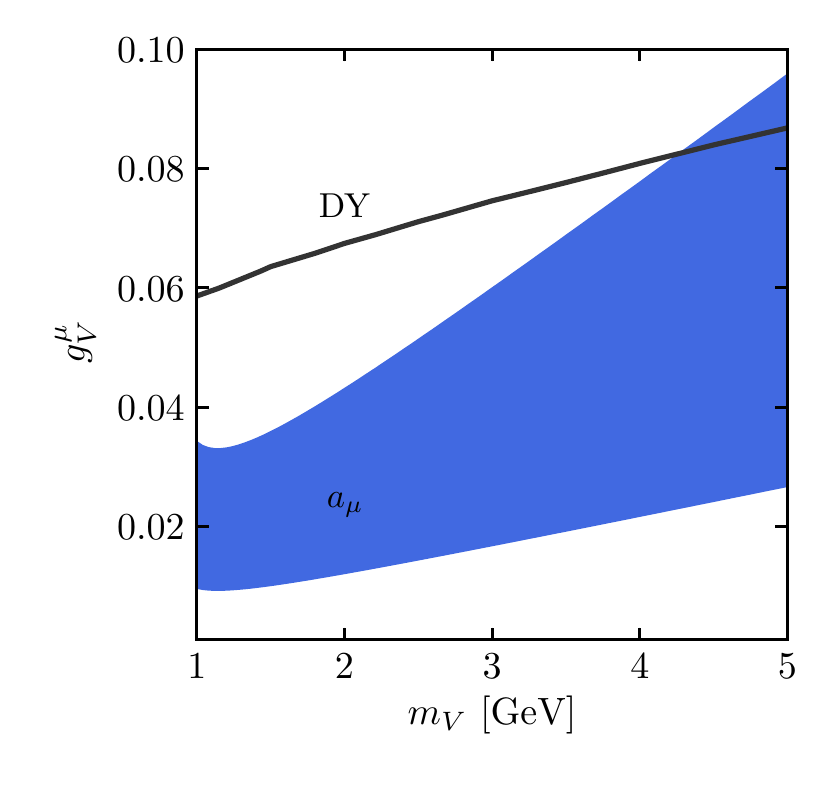}
\vspace{-4mm}
\caption{Upper panel: Constraints in the $g_V^\mu \hspace{0.25mm}$--$\hspace{0.25mm} g_A^\mu$ plane. The shown results were obtained for $m_V = 2.5 \, {\rm GeV}$ and $g_L^{sb} = -1.5 \cdot 10^{-8}$.  Lower panel: Constraints in the $m_V \hspace{0.25mm}$--$\hspace{0.25mm} g_V^\mu$ plane assuming the coupling relation  $g_A^\mu = 0.41 \hspace{0.5mm}  g_V^\mu$.  Consult the main text for further explanations.}
\label{fig:limits}
\end{center}
\end{figure}

The upper panel in Figure~\ref{fig:limits} compares the 95\%~CL constraint in the $g_V^\mu \hspace{0.25mm}$--$\hspace{0.25mm} g_A^\mu$ plane that derives from our fit to the $Z$-boson line shape for $m_V = 2.5 \, {\rm GeV}$ (black) to the regions preferred by $P_5^\prime$~(green), $R_K$~(yellow) and $R_{K^\ast}$~(red) and $a_\mu$~(blue).  The parameter space above and to the right of the black curve is excluded. In the case of the flavour observables the favoured parameter space corresponds to the $\Delta \chi^2 = 4$ regions obtained in~\cite{Sala:2017ihs} for $g_L^{sb} = -1.5 \cdot 10^{-8}$, while in the case of $a_\mu$ we have employed the $3 \sigma$ bound $\Delta a_\mu \in [49, 527] \cdot 10^{-11}$~\cite{gmtwo}. From the panel it is evident that the model-independent constraint that arises from the DY data excludes parts of the parameter space favoured by the $b \to s \ell^+ \ell^-$ anomalies. In particular, coupling choices that accommodate the deviation seen in $P_5^\prime$ are constrained. We now focus on the region of the $g_V^\mu \hspace{0.25mm}$--$\hspace{0.25mm} g_A^\mu$ plane in which the discrepancy between SM and data for $a_\mu$ is improved by the one-loop corrections due to the exchange of a light di-muon spin-1 resonance~(cf.~\cite{Jegerlehner:2009ry})
\begin{equation}
\Delta a_\mu = \frac{\left (g_V^\mu \right )^2 - 5 \left (g_A^\mu \right )^2}{12 \pi^2}  \frac{m_\mu^2}{m_V^2} + {\cal O} \left ( m_\mu^4/m_V^4 \right ) \,.
\end{equation}
In this region, we observe that our new constraint disfavours most of the parameter space that provides a simultaneous explanation of $P_5^\prime$, $R_K$, $R_{K^\ast}$ and $a_\mu$. Given the weak mass dependence of~(\ref{eq:masterformula}), we expect this conclusion to hold in the full range $m_V \in [2, 3] \, {\rm GeV}$ of spin-1 resonance interpretations of the flavour anomalies.

In the lower panel of Figure~\ref{fig:limits}, we compare the 95\%~CL bound in the $m_V \hspace{0.25mm}$--$\hspace{0.5mm} g_V^\mu$ plane following from measurements of the $m_{\mu \mu}$ spectrum in DY production~(black) to the region favoured by $a_\mu$~(blue). The shown results have been obtained for $g_A^\mu = 0.41 \hspace{0.5mm} g_V^\mu$. We see that our new DY constraint shrinks the allowed parameter space for such fine-tuned solutions of the $a_\mu$ anomaly for resonances heavier than about~$4.2 \, {\rm GeV}$. Spin-1 resonance explanations of~$a_\mu$ that do not rely on a cancellation in the combination $\left (g_V^\mu \right )^2 - 5 \left (g_A^\mu \right )^2$ of couplings, such as solutions with $g_V^\mu \neq 0$ and $g_A^\mu = 0$, on the other hand, cannot be probed through $Z$-boson line shape measurements at present.

{\bf Conclusions.} The main goal of this letter was to point out that precision measurements of DY production provide sensitive probes of light di-leptonic resonances. In view of the various deviations from SM predictions observed in rare semi-leptonic $B$ decays, we have applied our general observation to the case of GeV-mass di-muon spin-1 resonances. Specifically, we have analysed the distortions that FSR of such mediators imprints on the di-muon invariant mass spectrum as measured in $pp \to Z/\gamma^\ast \to \mu^+ \mu^-$ at the LHC. For simplified-model realisations that allow one to qualitatively reproduce the $P_5^\prime$, $R_K$, $R_{K^\ast}$ and $a_\mu$ anomalies, we have found that the $Z$-boson line shape develops a pronounced radiative tail that amounts to a relative enhancement of ${\cal O} (5\%)$ at $m_{\mu \mu} \simeq 75 \, {\rm GeV}$ compared to the SM prediction.

Motivated by this finding we have derived model-independent bounds on the muon couplings of spin-1 mediators using DY data from LHC Run II. Our analysis shows that the existing precision measurements of DY production  put non-trivial constraints on the parameter space of light di-muon resonance models~\cite{Sala:2017ihs} that aim at explaining the tensions seen in rare semi-leptonic $B$ decays. In particular, they disfavour almost all model realisations that can simultaneously accommodate the $P_5^\prime$, $R_K$, $R_{K^\ast}$ and $a_\mu$ anomalies. Considering $a_\mu$ alone, we found instead that present $Z$-boson line shape fits can only probe fine-tuned GeV-mass explanations of the anomaly with $|g_A^\mu| \simeq 0.44 \hspace{0.5mm} |g_V^\mu|$.  Since the data set used to derive the constraints contains only $3.2 \, {\rm fb}^{-1}$ of integrated luminosity collected at $\sqrt{s} = 13 \, {\rm TeV}$, future analyses performed at LHC~Run~II and beyond are expected to strengthen the obtained bounds in case no deviations from the $m_{\mu \mu}$ spectrum as predicted in the SM are found. 

While in our work we have focused our attention  on light di-muon spin-1 resonances, precision measurements of the  kinematic distributions of the final-state leptons in $pp \to Z/\gamma^\ast \to \ell^+ \ell^-$  can also be used to search for and to constrain mediators preferentially coupling to electron pairs and/or of  different spin. A dedicated study of the DY constraints on alternative light di-lepton resonance scenarios, while beyond the scope of this letter, thus seems to be a worthwhile exercise. 

\begin{acknowledgments} 
{\bf Acknowledgements.}  We thank David Straub for useful discussions concerning~\cite{Sala:2017ihs}. The work of FB is supported by the Science and Technology Facilities Council (STFC). UH acknowledges the support of the CERN Theoretical Physics Department. 
\end{acknowledgments}

\end{document}